\documentstyle[amssymb,aps,preprint,12pt]{revtex}

\begin{document}
\title{Electron transport in a quasi-one dimensional channel on suspended
helium films}
\author{Sviatoslav S. Sokolov}
\address{Departamento de F\'{\i}sica, Universidade Federal de S\~{a}o\\
Carlos, 13565-905, S\~{a}o Carlos, S\~{a}o Paulo, Brazil and B. I. Verkin\\
Institute for Low Temperature Physics and Engineering, National Academy of\\
Sciences of Ukraine, 61103, Kharkov, Ukraine}
\author{Nelson Studart}
\address{Departamento de F\'{\i}sica, Universidade Federal de S\~{a}o\\
Carlos, 13565-905, S\~{a}o Carlos, S\~{a}o Paulo, Brazil}
\date{\today}
\maketitle

\begin{abstract}
Quasi-one dimensional electron systems have been created using a suspended
helium film on a structured substrate. The electron mobility along the
channel is calculated by taking into account the essential scattering
processes of electrons by helium atoms in the vapor phase, ripplons, and
surface defects of the film substrate. It is shown that the last scattering
mechanism may dominate the electron mobility in the low temperature limit
changing drastically the temperature dependence of the mobility in
comparison with that controlled by the electron-ripplon scattering.
\end{abstract}


There has been great interest in the study of the quasi-one-dimensional
electron system (Q1DES) produced on the liquid helium surface\cite%
{kovnik92,kirmonkov93,kovnikyayam98,kovnikglad98,haracrfoz98,valkerheijden98,valkering98,gladnikkov00,klierdoicleid00}
in which the motion of usual quasi-two-dimensional (Q2D) surface electrons
(SE)\cite{review} is restricted in one more spatial direction. In most
methods to realize the Q1D electron system, the charged channels are formed
when parallel strips of a dielectric substrate are filled with liquid helium
due to the capillary forces. The curvature radius $R$ of the liquid into the
strip can be varied in wide range. The profile of the liquid surface across
the channel ($y$ direction) can be assumed to have a semicircular form $%
z_{0}=R\left[ 1-\sqrt{1-\left( y/R\right) ^{2}}\right] \simeq y^{2}/2R$ for $%
y\ll R$, where $R=\alpha /\rho gH$ with $\alpha $ and $\rho $ are the
surface tension and the helium density, respectively, $g$ is the
acceleration of the gravity, and $H$ is the bulk helium level below the
stripped structure. The electron confinement across the channel is achieved
by the action of a holding electric field $E_{\perp }$ along the normal
direction to the liquid surface ($z$ axis). As a result the electron near
the channel bottom is subjected to the parabolic potential $U(y)=eE_{\perp
}z_{0}\simeq m\omega _{0}^{2}y^{2}/2,$ where $\omega _{0}=(eE_{\perp
}/mR)^{1/2}$.\cite{kovmon86}

Quite recently, it was shown that very stable suspended helium films over a
structured substrate can be created with arbitrary thickness, as depicted in
Fig. 1.\cite{valkering98,valkering00} Under certain conditions (the distance
between the elevations must be smaller than the capillary length), the
helium film does not follow the substrate form but fills the depressions
through capillary condensation. The film thickness $d$ is controlled by
channel width $W$ and the bulk level height $H$. This method is useful to
laterally confine the electrons in a sub-micrometer case and generate Q1DES
above helium films. This system can be modelled in a similar form as before,
but an effective holding field $E_{\perp }^{\ast }$ replaces now $E_{\perp }$
in the confined frequency $\omega _{0}$ due to the contribution of the
polarization interaction of the electron with the solid substrate: $E_{\perp
}^{\ast }=E_{\perp }+\left( \Lambda _{1}/e\right) \langle
1|(z+d)^{-2}|1\rangle =E_{\perp }+(2\Lambda _{1}\gamma ^{2}/e)f(\gamma d)$
where the angular brackets denote averaging over the electron wave function $%
|1\rangle =2\gamma ^{3/2}z\exp (-\gamma z)$ for the motion in the $z$
direction ($\gamma $ is the localization parameter \cite{sokrinostud97}), $%
\Lambda _{1}=e^{2}\varepsilon _{\text{He}}(\varepsilon _{s}-\varepsilon _{%
\text{He}})/\left[ (\varepsilon _{\text{He}}+1)^{2}(\varepsilon
_{s}+\varepsilon _{\text{He}})\right] $ with $\varepsilon _{\text{He}}$ and $%
\varepsilon _{s}$ the dielectric constants of liquid helium and substrate,
respectively; $d$ is film thickness near $y=0$, $f(x)=\left[
1+2x+4x(1+x)\exp (2x)\text{Ei}(-2x)\right] ,$ and Ei$(x)$ is integral
exponential function. According to our estimates polarization effects
becomes dominant for $d\lesssim 10^{-6}$ cm, especially for substrates with
large $\varepsilon _{s}$.

For wide enough rectangular strips, the curvature effects can be discarded
especially for small $d$ and high electron densities which lead to a flat
profile in the centre of channel. However, charged electrodes may be
arranged in such a way that the applied gate voltage produces an effective
lateral electrostatic confinement.\cite{valkerheijden98,klierdoicleid00} The
electrostatic potential should be calculated using the Poisson equation with
appropriate boundary conditions. However for $|y|\ll W$, the same model can
be used, but $U(y)$ now depends on $\omega _{\text{conf}}^{2}=\omega
_{0}^{2}+\omega _{\text{es}}^{2}$, where $\omega _{\text{es}}$ is the
characteristic electrostatic frequency.

In this paper, we study the transport properties of the Q1DES over helium
film. Despite of different possible ways to create the system, the electron
states inside the channel can be simply described. Indeed the electron
potential energy can be written as

\begin{equation}
V(y,z)=eE_{\perp }z-\frac{\Lambda _{0}}{z}-\frac{\Lambda _{1}}{z+d}+\frac{%
m\omega _{\text{conf}}^{2}y^{2}}{2}  \label{1}
\end{equation}%
where $\Lambda _{0}=e^{2}(\varepsilon _{\text{He}}-1)\left[ 4(\varepsilon _{%
\text{He}}+1)\right] $ and the characteristics of the channel geometry are
given by $\omega _{\text{conf}}$ within the harmonic approximation. The
energy spectrum and the wave function for electron in the plane are $%
E_{n,k_{x}}=\hslash ^{2}k_{x}^{2}/2m+\hslash \omega _{\text{conf}}\left(
n+1/2\right) $ and $\chi _{n}(x,y)=$[$\exp (ik_{x}x)\exp \left(
-y^{2}/2l^{2}\right) H_{n}\left( y/l\right) ]/(\pi
^{1/2}lL_{x}2^{n}n!)^{1/2} $ respectively, where $l=\left( \hslash /m\omega
_{\text{conf}}\right) ^{1/2} $ and $H_{n}(x)$ are the Hermite polynomials.
The parameter $l$ gives the scale of electron localization in the $y$
direction ($l$ yields $3.4\times 10^{-6}$ cm for $\omega _{\text{conf}%
}=10^{11}$ Hz). Typical values of $\omega _{0}$ are in the range $%
10^{10}-10^{11}$ Hz (for $E_{\perp }=1-3$ kV/cm and $R=5\times 10^{-4}$cm).%
\cite{sokhaistud95} Then the approximation $z_{0}\simeq y^{2}/R$ is rather
good and the condition $y\ll R$ is well satisfied.

The electron multisubband spectrum leads to rather interesting transport
properties along the channel. The main scattering mechanisms are the
electron interaction with atoms in the vapor phase predominating at $T>1$ K,
the electron-ripplon interaction at lower temperatures, and the electron
scattering by surface defects at the helium-substrate interface ($z=-d)$. As
it was shown the latter scattering can dominate the SE mobility over a
helium film in the low temperature regime.\cite{coimbrasokrinostud02} The
electron mobility in the Q1D channel over bulk helium was calculated in Ref. %
\cite{sokhaistud95} taking into account the population of the excited
subbands in the $y$ direction when $\hslash \omega _{\text{conf}}\lesssim T$
(note that with $\hslash \omega _{\text{conf}}\simeq 0.8$ K for $\omega _{%
\text{conf}}\simeq 10^{11}$ Hz). However the contribution of $n>1$ subbands
can be discarded for $T\ll \hslash \omega _{\text{conf}}$. Here we limit
ourselves to this regime and neglect quantum statistics effects (Fermi
energy is much smaller than $T$).

The electron mobility along the channel in the limit $\hslash \omega _{\text{%
conf}}/T\gg 1$ is given as\cite{sokhaistud95}

\begin{equation}
\mu =\frac{2}{\sqrt{\pi }}\frac{e}{m}\left( \frac{\hslash \omega _{\text{conf%
}}}{T}\right) ^{3/2}\int_{0}^{\infty }\frac{\sqrt{x}\exp \left( -\hslash
\omega _{\text{conf}}x/T\right) }{[\nu _{g}(x)+\nu _{r}(x)+\nu _{d}(x)]},
\label{5}
\end{equation}%
where $x=\hslash k_{x}^{2}/(2m\omega _{\text{conf}}).$ The collision
frequencies $\nu _{g}(x),$ $\nu _{r}(x),$ and $\nu _{d}(x)$ denote the
electron collisions with vapor atoms, ripplons, and defects, respectively.

The collision frequency with helium atoms is given by $\nu _{g}(x)=3\hslash
n_{g}\gamma A/8m\sqrt{x}$, where $n_{g}$ is the volume concentration of
helium atoms and $A\simeq 4.676\times 10^{-16}$ cm$^{-2}$ is the scattering
cross-section. Note that the contribution of $\nu _{g}(x)$ becomes
negligible for $T<1$ K because $n_{g}$ decays exponentially with $T$. The
electron mobility is dominated from contributions of $\nu _{r}(x)$ and $\nu
_{d}(x)$ in that temperature range.

In order to calculate $\nu _{r}(x),$ we use the expression\cite{sokhaistud95}

\begin{equation}
\nu _{r}(k_{x})=\frac{2\pi }{\hslash S}\sum_{{\bf q}}|\langle 0|\exp
(iq_{y}y)|0\rangle |^{2}|\langle 1|V_{r{\bf q}}(z)|1\rangle |^{2}(2N_{q}+1)%
\frac{q_{x}}{k_{x}}\delta \left( E_{k_{x}-q_{x}}-E_{k_{x}}\right) .
\label{6}
\end{equation}%
Here $E_{k_{x}}=\hslash ^{2}k_{x}^{2}/2m$, ${\bf q}$ is the 2D wave vector,
and $V_{r{\bf q}}(z)=\left[ \hslash q\tanh (qd)/2\rho \omega _{q}\right]
^{1/2}eE_{\perp }^{\ast }$ for a thin helium film, $\omega _{q}^{2}=[(\alpha
/\rho )q^{3}+g^{\prime }q]\tanh (qd)$ is ripplon dispersion law with $%
g^{\prime }=g+3n_{He}\beta /(\rho d^{4})$ where $\beta $ is the van der
Waals constant of the substrate, $n_{He}$ is volume concentration in liquid
helium.\cite{monshik82} Long wavelength ripplons do mainly contribute to the
scattering with $N_{q}\simeq 2T/\hslash \omega _{q}$. Straightforward
calculation of Eq. (\ref{6}) leads to

\begin{equation}
\nu _{r}(x)=\frac{e^{2}(E_{\perp }^{\ast })^{2}T}{4\alpha \hslash ^{2}\omega
_{\text{conf}}}\cdot \frac{\exp \left[ 4\left( x+x_{c}\right) \right] }{%
\sqrt{x^{2}+x_{c}x}}\left[ 1-%
\mathop{\rm erf}%
\left( 2\sqrt{x+x_{c}}\right) \right] .  \label{7}
\end{equation}%
Here $x_{c}=\hslash \rho g^{\prime }/8\alpha m\omega _{\text{conf}}.$ In the
limit $d\rightarrow \infty $ and guessing $g=0$ one obtains

\begin{equation}
\nu _{r}(x)=\frac{e^{2}(E_{\perp }^{\ast })^{2}T}{4\alpha \hslash ^{2}\omega
_{0}}\cdot \frac{\exp (4x)}{x}[1-%
\mathop{\rm erf}%
(2\sqrt{x})].  \label{8}
\end{equation}%
which reproduces the result of Ref. \cite{kovmon86}.

The general expression for $\nu _{d}(k_{x})$ can be written as 
\begin{equation}
\nu _{d}(k_{x})=\frac{2\pi }{\hslash S}\sum_{{\bf q}}|\langle 0|\exp
(iq_{y}y)|0\rangle |^{2}|\langle 1|V_{d{\bf q}}(z)|1\rangle |^{2}|\xi _{s%
{\bf q}}|^{2}\frac{q_{x}}{k_{x}}\delta \left(
E_{k_{x}-q_{x}}-E_{k_{x}}\right) ,  \label{9}
\end{equation}%
where $\xi _{s{\bf q}}$ is the Fourier transform of the solid interface
displacement from equilibrium position $z=-d$, $V_{d{\bf q}}(z)=-\Lambda
_{1}qK_{1}[q(z+d)]/(z+d)$, with $K_{1}(x)$ the modified Bessel function.\cite%
{coimbrasokrinostud02} To calculate the electron-defect contribution, we use
the well-known Gaussian two-parameter model for surface defects in which the
correlation function $\langle \xi ({\bf r})\xi ({\bf r}^{\prime })\rangle
=\xi _{0}^{2}\exp \left[ -|{\bf r-r}^{\prime }|^{2}/a^{2}\right] $ depends
on $\xi _{0}$ and $a$ playing the roles of characteristic defect height and
width respectively.\cite{prange78} This model leads to $\langle |\xi _{s{\bf %
q}}|^{2}\rangle $ $=\pi \xi _{0}^{2}a^{2}\exp \left( -q^{2}a^{2}/4\right) $
and was used to explain the SE transport properties over a thin helium film%
\cite{coimbrasokrinostud02} and over solid hydrogen.\cite{sokrinostud95} The
final expression for $\nu _{d}(x)$ is 
\begin{equation}
\nu _{d}(x)=\frac{32m\Lambda _{1}^{2}\xi _{0}^{2}a^{2}\exp \left(
-2a^{2}x/l^{2}\right) }{\hslash ^{3}l^{4}x^{1/2}}\Phi _{d}(x)  \label{10}
\end{equation}%
where 
\[
\Phi _{d}(x)=\int_{0}^{\infty }\frac{dy}{\sqrt{y}}(x+y)^{2}\exp \left[
-\left( 4+\frac{2a^{2}}{l^{2}}\right) y\right] \varphi _{d}^{2}\left( \frac{%
\sqrt{2(x+y)}}{\gamma l}\right) 
\]%
and 
\[
\varphi _{d}(x)=\frac{\exp (2\gamma d)}{x}\int_{2\gamma d}^{\infty }\frac{ds%
}{s}(s-2\gamma d)^{2}K_{1}(xs)\exp (-s). 
\]

For a ideal surface substrate ($\nu _{d}(x)=0$) the electron mobility is
determined by $\nu _{r}(x)$ and is given by

\begin{equation}
\mu _{r}\simeq 6\mu _{\perp }\left[ 1+\frac{32}{3\pi }\left( \frac{T}{%
\hslash \omega _{\text{conf}}}\right) ^{1/2}\right]  \label{11}
\end{equation}%
for large enough $d$ and $x_{c}$ so small that the condition $T\gg \hslash
\omega _{\text{conf}}x_{c}$ is fulfilled, even though $T\ll \hslash \omega _{%
\text{conf}}$. Here $\mu _{\perp }=\alpha \hslash /[em(E_{\perp }^{\ast
})^{2}]$. In case of thin films with $T\ll \hslash \omega _{\text{conf}%
}x_{c} $, the asymptotic expression for ripplon limited mobility is%
\begin{equation}
\mu _{r}\simeq \frac{8\mu _{\perp }}{\sqrt{\pi }}\left( \frac{\hslash \omega
_{0}x_{c}}{T}\right) ^{1/2}\frac{\exp \left( -4x_{c}\right) }{1-%
\mathop{\rm erf}%
\left( 2\sqrt{x_{c}}\right) }.  \label{12}
\end{equation}%
By comparing Eqs. (\ref{11}) and (\ref{12}), one can see the drastic change
in the temperature dependence of the mobility from a thick to a thin film.
One estimates this transition at $d\lesssim 10^{-6}$ cm for actual substrate
materials.

The situation becomes rather interesting when one includes the defect
contributions of solid substrates. Indeed for realistic values of $a=10^{-6}$
cm, $d$ in the same range, and $\xi _{0}=10^{-7}$ cm, $\nu _{d}(x)$ is near
two orders of magnitude larger than $\nu _{r}(x)$ for $x=x_{T}=T/\hslash
\omega _{\text{conf}}$ and gives the major contribution to the integral of
Eq. (\ref{5}). In such a condition, the defect-limited mobility is given as 
\begin{equation}
\mu _{d}\simeq \frac{e}{m\nu _{d}^{(0)}}\left( \frac{T}{\hslash \omega _{%
\text{conf}}}\right) ^{1/2}  \label{13}
\end{equation}%
valid for above-mentioned values of $\xi _{0}$ and $a$ and $T\gtrsim 0.1$ K.
Here $\nu _{d}^{(0)}=\pi m\Lambda _{1}^{2}\xi _{0}^{2}a^{2}\gamma
^{4}f^{2}(2\gamma d)/\hslash ^{3}$.

The results for mobility, given by Eqs. (\ref{11}-\ref{13}), are obtained in
the single-electron approximation (SEA). Meantime, by increasing electron
density, the effects of electron correlations can influence the electron
transport properties. To take these effects into account one can apply the
so-called complete control approach (CCA) or the Boltzmann shifted
distribution approximation, where we assume that the electrons have equal
drift velocity $u$ and their distribution function is close to exp$\left[
-\hslash ^{2}k_{x}^{2}/2mT+\hslash k_{x}u/T\right] $. This approach has been
successfully used in calculating the electron mobility in both Q1DES and
Q2DES over bulk helium.\cite{sokhaistud95,bungrigkir90} It is valid when the
electron-electron collision frequency is significantly larger than the
frequencies $\nu _{r}(x)$, $\nu _{g}(x)$, and $\nu _{d}(x)$. The method of
calculation is described in details in Ref. \cite{sokhaistud95}. The final
expression for the ripplon-limited mobility in the CCA is 
\begin{equation}
\mu _{r}^{\text{(cca)}}\simeq 2\mu _{\perp }\left[ 1+\frac{4}{\pi }\left( 
\frac{T}{\hslash \omega _{\text{conf}}}\right) ^{1/2}\right]  \label{14}
\end{equation}%
for thick films where $T\gg \hslash \omega _{\text{conf}}x_{c}$. In the
opposite limit $\mu _{r}^{\text{(cca)}}=(\pi /4)\mu _{r}$, for $T\ll \hslash
\omega _{\text{conf}}x_{c}$ where $\mu _{r}$ is given by Eq. (\ref{12}).
Comparing with Eqs. (\ref{11}) and (\ref{12}), one concludes that CCA gives
the same qualitatively dependences of the ripplon-limited mobility on
temperature and effective holding field as SEA. However the absolute values
of mobility are smaller in the CCA showing the influence of electron
correlations on transport in Q1DES. The defect-limited mobility in CCA is $%
\mu _{d}^{\text{(cca)}}=\mu _{d}/4$ where $\mu _{d}$ is given by Eq. (\ref%
{13}).

In conclusion we have investigated theoretically the properties of Q1DES
over suspended helium films. Film effects modify the confinement potential
across the channel and the electron mobility at low temperatures is limited
by ripplon scattering and mainly by surface defects at the helium film
substrate interface. The latter scattering is dominant for thin films with $%
d\sim 10^{-6}$ cm and leads to the increase of electron mobility with
temperature whereas the ripplon-limited mobility should decrease in this
limit. Such a prediction can be tested in experimental attempts to observe
the influence of different scattering mechanisms in the electron transport
of Q1DES over helium film for temperatures below $1$ K.

This work was supported by the Funda\c{c}\~{a}o de Amparo \`{a} Pesquisa do
Estado de S\~{a}o Paulo (FAPESP)\ and the Conselho Nacional de
Desenvolvimento Cient\'{\i}fico e Tecnol\'{o}gico (CNPq). The authors are
indebted to Professor F. M. Peeters for discussion of the results.

\bigskip

\bigskip

\begin{center}
{\bf FIGURE CAPTION}
\end{center}

Fig. 1. Schematic diagram of a suspended film on a structured substrate.


\begin{references}
\bibitem{kovnik92} Yu. Z. Kovdrya and V. A. Nikolaenko, Fiz. Nizk. Temp. 
{\bf 18}, 1278 (1998) [Low Temp. Phys. {\bf 18}, 894 (1992)].

\bibitem{kirmonkov93} O. I. Kirichek, Yu. P. Monarkha, Yu. Z. Kovdrya, and
V. N. Grigor'ev, Fiz. Nizk. Temp. {\bf 19}, 458 (1993) [Low Temp. Phys. {\bf %
19}, 323 (1993)].

\bibitem{kovnikyayam98} Yu. Z. Kovdrya, V. A. Nikolaenko, H. Yayama, A.
Tomokiyo, O. I. Kirichek, and O. B. Berkutov, J. Low Temp. Phys.{\bf \ 110},
191 (1998).

\bibitem{kovnikglad98} Yu. Z. Kovdrya, V. A. Nikolaenko, S. P. Gladchenko,
and S. S. Sokolov, Fiz. Nizk. Temp.{\bf \ 24}, 1113 (1998) [Low Temp. Phys. 
{\bf 24}, 837 (1998)].

\bibitem{haracrfoz98} R. J. F. van Haren, G. Acres, P. Fozooni, A.
Kristensen, M. J. Lea, P. J. Richardson, A. M. C. Valkering, and R. W. van
der Heijden, Physica B {\bf 251}, 656 (1998).

\bibitem{valkerheijden98} A. M. C. Valkering and R. W. van der Heijden,
Physica B, {\bf 251}, 652 (1998).

\bibitem{valkering98} A. M. C. Valkering, {\it Surface electrons in
restricted geometry}, PhD thesis, Eindhoven (1998), unpublished.

\bibitem{gladnikkov00} S. P. Gladchenko, V. A. Nikolaenko, Yu. Z. Kovdrya,
and S. S. Sokolov, Fiz. Nizk. Temp. {\bf 27}, 3 (2000) [Low Temp. Phys. {\bf %
27}, 3 (2000)].

\bibitem{klierdoicleid00} J. Klier, I. Doicescu, and Paul Leiderer, J. Low
Temp. Phys. {\bf 121}, 603 (2000).

\bibitem{review} See for a review, {\it Two-dimensional electron systems in
helium and other substrates, }edited by E. Y. Andrei (Kluwer, Dordrecht,
1997).

\bibitem{kovmon86} Yu. Z. Kovdrya and Yu. P. Monarkha, Fiz. Nizk. Temp. {\bf %
12}, 1011 (1986) [Low Temp. Phys. {\bf 12}, 571 (1986)].

\bibitem{valkering00} A. Valkering, J. Klier, and P. Leiderer, Physica B 
{\bf 284}, 172 (2000). See also D. Marty, J. Phys. C: Solid State Phys. {\bf %
19}, 6097 (1986) for earlier attempts.

\bibitem{sokrinostud97} S. S. Sokolov, J.-P. Rino, and N. Studart, Phys.
Rev. B{\bf \ 55,} 14473 (1997).

\bibitem{sokhaistud95} S. S. Sokolov, G. Q. Hai, and N. Studart, Phys. Rev.
B {\bf 51}, 5977 (1995).

\bibitem{coimbrasokrinostud02} D. Coimbra, S. S. Sokolov, J. P. Rino, and N.
Studart, to appear in J. Low Temp. Phys..

\bibitem{monshik82} Yu. P. Monarkha and V. B. Shikin, Fiz. Nizk. Temp. {\bf 8%
}, 563 (1982) [Sov. J. Low Temp. Phys. {\bf 8}, 279, (1982)].

\bibitem{prange78} R. E. Prange and T. W. Nee, Phys. Rev. {\bf 168}, 779
(1978).

\bibitem{sokrinostud95} S. S. Sokolov, J.-P. Rino, and N. Studart, Phys.
Rev. B{\bf \ 51,} 11068 (1995).

\bibitem{bungrigkir90} V. A. Buntar', V. N. Grigoriev, O. I. Kirichek, Yu.
Z. Kovdrya, Yu. P. Monarkha, and S. S. Sokolov, J. Low Temp. Phys. {\bf 79},
323 (1990).
\end{references}
\end{document}